# Effect of carbon nanotube doping on critical current density of MgB$_2$ superconductor

S.X. Dou, W.K. Yeoh, J. Horvat and M. Ionescu
*Institute for Superconducting and Electronic Materials, University of Wollongong, Northfields Avenue, Wollongong, NSW 2522, Australia*

The effect of doping MgB$_2$ with carbon nanotubes on transition temperature, lattice parameters, critical current density and flux pinning was studied for MgB$_{2-x}$C$_x$ with x = 0, 0.05, 0.1, 0.2 and 0.3. The carbon substitution for B was found to enhance J$_c$ in magnetic fields but depress T$_c$. The depression of T$_c$, which is caused by the carbon substitution for B, increases with increasing doping level, sintering temperature and duration. By controlling the extent of the substitution and addition of carbon nanotubes we can achieve the optimal improvement on critical current density and flux pinning in magnetic fields while maintaining the minimum reduction in T$_c$. Under these conditions, J$_c$ was enhanced by two orders of magnitude at 8T and 5K and 7T and 10K. J$_c$ was more than 10,000A/cm$^2$ at 20K and 4T and 5K and 8.5T, respectively.

The effect of C-doping on superconductivity in MgB$_2$ compound has been studied by several groups [1-8]. The results on C solubility and the effect of C-doping on T$_c$ reported so far vary significantly due to the precursor materials, fabrication techniques and processing conditions used. It appears that lower sintering temperatures and short sintering times result in an incomplete reaction and hence lower C solubility in MgB$_2$. Ribeiro et al. used Mg and B$_4$C as precursors to synthesize C doped MgB$_2$ by sintering at 1200°C for 24 hours [9]. A neutron diffraction study confirmed that the most likely solubility of C in MgB$_2$ is up to around 10% of C in the boron sites, resulting in a large drop of both T$_c$ and the *a*-axis lattice parameter [10]. Recently, Lee et al. synthesised C-doped single crystalline MgB$_2$ at high pressure (5GPa) and high temperature (1600°C), obtaining the C solubility of 15% at the boron sites and T$_c$ depression to below 3K [11]. All these studies on C doping into MgB$_2$ have only focused on the effect on superconductivity. From the applications point of view, the effect of C doping on the flux pinning properties is crucially important. The author's group has reported a significant improvement in J$_c$(H) and H$_{irr}$ in MgB$_2$ through nano-SiC doping [12]. Recently, the effects of C doping on the flux pinning and critical current density in MgB$_2$ has been studied using amorphous carbon [13] and diamond [14], both showing improvement of J$_c$ at elevated magnetic fields. Wei et al. have studied the superconductivity of MgB$_2$-carbon nanotube composites [15]. However, the effect of carbon nanotube doping on critical current density and flux pinning has not been reported. Among various carbon precursors, carbon nanotubes are particularly interesting as their special geometry (high aspect ratio and nanometer diameter) may induce more effective pinning centres compared to other carbon-containing precursors. In this letter we report the results on control of the extent of carbon nanotube substitution and addition to achieve an enhancement of critical current density and flux pinning by two orders of magnitude in magnetic fields.

Polycrystalline samples of MgB$_{2-x}$C$_x$ were prepared through a reaction in-situ process [16, 17]. High purity powders of magnesium (99%), amorphous boron (99%) and multi-walled carbon nanotubes of 20-30nm diameter were weighed out according to the nominal atomic ratio of MgB$_{2-x}$C$_x$ with x = 0, 0.05, 0.1, 0.2, 0.3, and well-mixed through grinding. We stress that the values for x are the nominal values throughout the paper, and actual substitution of C for B will be shown to be less than these values. The powders were pressed into pellets of 10 mm in diameter and 3 mm in thickness using a hydraulic press. The pellets were sealed in Fe tubes, then heat treated at 700 to 1000°C for 10 to 120min in flowing high purity Ar. This was followed by a furnace cooling to room temperature. An un-doped sample was also made under the same conditions for the use as a reference sample. The phase and crystal structure of all the samples was obtained from X-ray diffraction (XRD) patterns using a MAC Science MX03 diffractometer with Cu $K\alpha$ radiation. Si powder was used as an internal standard to calculate the lattice parameters. The grain morphology and microstructure were also examined by scanning electron microscope (SEM) and transmission electron microscope (TEM).

The magnetization was measured over a temperature range of 5 to 30 K using a Physical Property Measurement System (PPMS, Quantum Design) in a time-varying magnetic field with sweep rate 50 Oe/s and amplitude 8.5T. Bar shaped samples with a size of 4 x 3 x 0.5 mm$^3$ were cut from each pellet for magnetic measurements. The magnetic measurements were performed by applying the magnetic field parallel to the longest sample axis. The magnetic $J_c$ was calculated from the height of the magnetization loop $\Delta M$ using the Bean model: $J_c=20 \ \Delta M/[a/(1-a^2/3b)]$, with a and b as the dimensions of the sample perpendicular to the direction of applied magnetic field and a < b. $J_c$ versus magnetic field has been measured up to 8.5 T. The low field $J_c$ below 10 K could not be measured due to flux jumping. The $T_c$ was determined by measuring the real part of the ac susceptibility at a frequency of 117 Hz and an external magnetic field of 0.1 Oe. $T_c$ was defined as the onset of the diamagnetism.

Fig. 1 shows lattice parameter *a,* and unit cell volume vs. sintering temperature for sample doped at x = 0.2. The *a*-axis decreases monotonically with increasing sintering temperature. The *c*-axis varies very little with the sintering temperature and consequently the volume changes in a similar manner as *a* (Fig.1). The decrease of *a*-axis is an indication of the carbon substitution for boron. The decrease of *a*-axis is more pronounced at temperatures above 900°C, because of the enhanced carbon substitution at these temperatures, which is consistent with several recent papers [10, 11]. However, the substitution reaction in the present work is far from completion even at sintering temperature of 1000°C, in comparison with those treated at a higher temperature (1600°C) and high pressure [11]. Thus, we achieved a condition of partial substitution of C for B and addition of majority of C, which may react with B to form BC, as detected using EELS [18], or stay as carbon nanotubes. Inset to Fig.1 shows the XRD data for carbon nanotube doped samples, sintered at three different temperatures. The sample was a well developed MgB$_2$ phase, with only a small amount of MgO present, similar to other high-quality MgB$_2$ samples[12, 13, 17].

Fig. 2 shows the transition temperature ($T_c$) for the doped and undoped samples determined by ac susceptibility measurements. The $T_c$ onset for the undoped sample is ~ 38 K. For the sample doped at x=0.2 (10% of B) and sintered for a fixed period of 30min, the $T_c$ decreases with increasing sintering temperature. $T_c$ reaches 31K for the sintering temperature of 1000°C. This indicates that the extent of C substitution reaction increases with increasing temperature, resulting in $T_c$ depression, which is consistent with the recent reports [10, 11]. The effect of doping level on $T_c$ was also studied at fixed sintering temperature of 800°C for 30min. For this sintering temperature, $T_c$ only drops slightly, 2.0K at a C doping level of x = 0.3 (15% C doping). These results suggest that only a small amount of C nanotube powder was substituted in the B position in the samples sintered at low temperature and a short period, consistent with small crystal lattice contraction. In order to improve the $J_c$ at higher temperature, such as 20K, it is essential to maintain high $T_c$. The above results indicate that by manipulating the processing parameters we could control the $T_c$ while achieving a high level of C inclusion into MgB$_2$ sample, up to 10% of B. Because such C inclusion has little effect on $T_c$, the partial substitution of carbon for boron and partial addition of nanocarbon particles into MgB$_2$ matrix may enhance flux pinning within a wide range of temperatures.

Fig. 3 shows the $J_c(H)$ curves at 5K and 20K for the samples of MgB$_{2-x}$C$_x$, where x =

0, 0.05, 0.1, 0.2 and 0.3 are the nominal values for C content, with all the samples sintered at 800°C for 30min. All the $J_c(H)$ curves for doped samples have a higher $J_c$ than the undoped sample at high fields. The sample doped with 10% of carbon nanotubes (x = 0.2) gives the best $J_c$ at high fields: $J_c$ increases by a factor of 45 at 5K for the field of 8T, and at 20K for the field of 5T, as compared to the undoped sample. At higher doping level (x = 0.3), although the $J_c$ in low field regime was depressed, the rate of $J_c$ drop is much slower than for all other samples, clearly indicating strong flux pinning induced by the C nanotube doping.

Fig. 4 shows the $J_c(H)$ curves at 5 K and 20K for $MgB_{1.8}C_{0.2}$ sample sintered at temperatures from 700-1000°C for 30min. For comparison, the $J_c(H)$ curve for the undoped sample, $MgB_2$ sintered at 800°C is included. It is noted that the sintering temperature has a significant effect on the $J_c$ performance in the field. A general trend is such that the $J_c(H)$ characteristic is improved with increasing sintering temperature. Although the sample sintered at 1000°C has lower $J_c$ values in low field regime, its $J_c(H)$ curve crosses over the $J_c(H)$ for the other samples in higher fields. As higher sintering temperature promotes the C substitution reaction for B, the improved filed dependence of $J_c$ measured at lower temperatures is clearly attributable to the C substitution. However, because C substitution depresses $T_c$, the $J_c(H)$ behaviour for samples processed at high temperatures deteriorates above 20K. Thus, it is important to control the extent of C nanotube substitution and addition to achieve the best combination of the substitution induced flux pinning and C nanotube additive pinning.

Fig. 5 compares the $J_c(H)$ for carbon nanotubes and nano-C particle[13] doped $MgB_2$ at 5K and 20K. It is noted that the carbon nanotubes produced a stronger enhancement of $J_c$ than the nano-C particles. The reasons for C nanotube doping being far better than C nanoparticle doping can be explained as follows. The optimum doping level for nano-C particles is x = 0.05 (2.5% of B) while this level increases to x = 0.2 (10% of B) for C nanotube doping. This indicates that superconductivity of $MgB_2$ shows a higher tolerance to C nanotubes than C nanoparticles. Consequently, there can be higher concentration of nano-inclusions in the C nanotube doped sample than in the C nanoparticle doped ones, for the same value of $T_c$. Furthermore, the special geometry of C nanotubes is desirable for effective pinning.

In summary, the effect of C nanotube doping on lattice parameters, $T_c$, $J_c$ and flux pinning in $MgB_2$ was investigated under a wide range of processing conditions. It was found that substitution of C nanotube for B enhances the flux pinning but depresses $T_c$. By controlling the processing parameters an optimised $J_c(H)$ performance is achieved under a partial C substitution and C nano-addition. Under these conditions, $J_c$ was enhanced by two orders of magnitude at 8T and 5K , and at 7T and 10K. The $J_c$ was more than 10,000A/cm$^2$ at 20K in field of 4T and at 5K in field of 8.5T, respectively. Carbon nanotube inclusions and C substitution for B are proposed to be responsible for the enhancement of flux pinning in high fields.

Acknowledgment

The authors thank S. Soltanian and S.H. Zhou for their help in the experiment. This work was supported by the Australian Research Council, Hyper Tech Research Inc. OH USA, Alphatech International Ltd., NZ and the University of Wollongong. W. K. Yeoh received an Australia-Asia Award funded by the Australian Government.

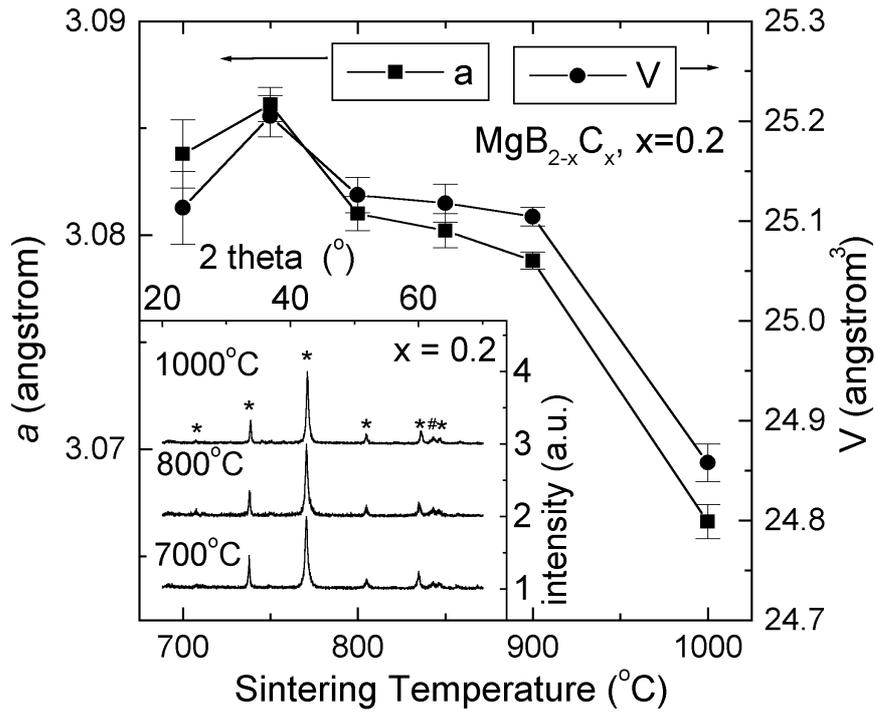

Figure 1: Variation of lattice parameter $a$ and unit cell volume of $MgB_{2-x}C_x$, with nominal x = 0.2, with sintering temperature. The carbon was in the form of multi-walled carbon nanotubes. Inset: XRD pattern for carbon nanotube doped $MgB_2$, sintered at temperatures as indicated in the Figure. Symbols * and # indicate the XRD peaks for $MgB_2$ and MgO, respectively.

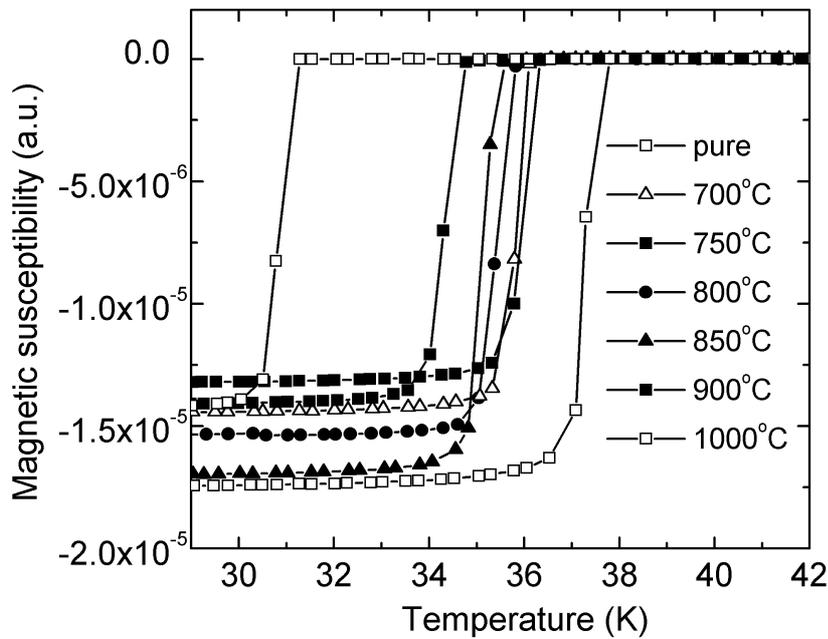

Figure 2. Magnetic AC susceptibility as a function of temperature for $MgB_{2-x}C_x$ sintered at different temperatures for 30 minutes. Carbon added was in the form of multi-walled carbon nanotubes.

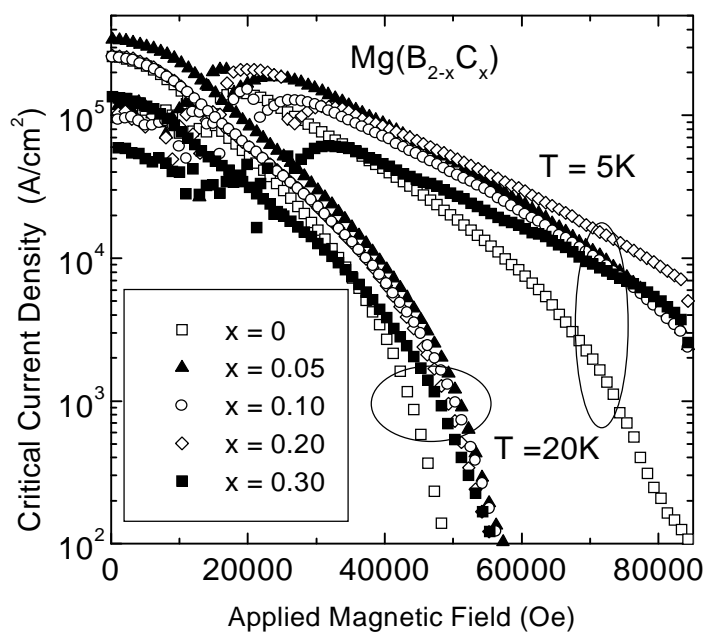

Figure 3. Critical current density as a function of magnetic field at 5K and 20K for different doping level of multi-walled carbon nanotubes.

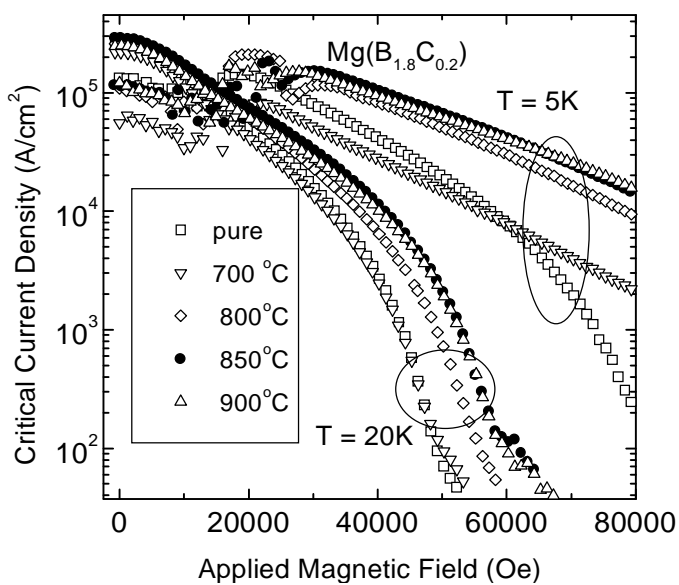

Figure 4. Critical current density as a function of magnetic field at 5K and 20K for $MgB_{1.8}C_{0.2}$, sintered at different temperatures for 30 minutes. The carbon was in the form of multi-walled carbon nanotubes.

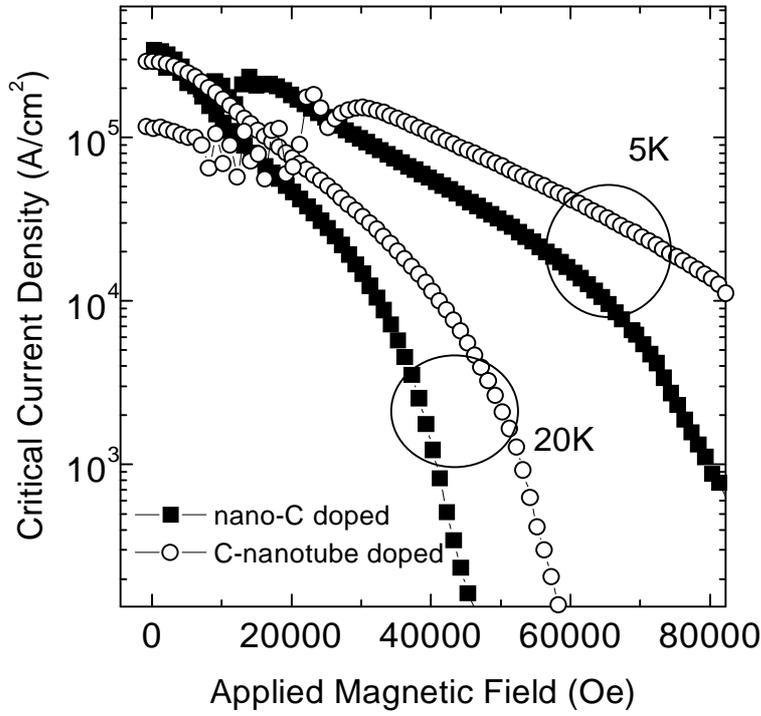

Figure 5: Field dependence of $J_c$ for 10 % nano-C doped (solid symbols) and 10% carbon nanotube doped (open symbols) $MgB_2$ at 5 and 20K. This doping level corresponds to the nominal x = 0.2. The sweep rate of the field was 50 Oe/s.